\documentclass[twoside,fleqn]{article}

\usepackage{espcrc2,feynmp,epsfig}

\newcommand{\psla}{p\kern-1.0ex/}

\def\beq{\begin{eqnarray}}
\def\eeq{\end{eqnarray}}

\newcommand{\AmS}{{\protect\the\textfont2
  A\kern-.1667em\lower.5ex\hbox{M}\kern-.125emS}}

\title{
\thispagestyle{empty}
\vspace{-3.5em}
\begin{flushright}
\normalsize
hep-ph/9907226\\
WUB 99-17
\end{flushright}
Phenomenology of $\eta\gamma$ and $\eta'\gamma$ transition
form factors\\ at large momentum transfer}

\author{Th.\ Feldmann\address{Fachbereich Physik, Universit\"at Wuppertal,
42097 Wuppertal, Germany}%
        \thanks{Supported by {\it Deutsche Forschungsgemeinschaft}}}
       
\begin{document}

\begin{abstract}
I discuss the progress in our theoretical understanding of
the $\eta\gamma$ and $\eta'\gamma$ transition form factors,
including the recent data from CLEO and L3
at large momentum transfer, $Q^2$.
The experimental data above $Q^2=1$~GeV$^2$ can be well described by the
hard scattering approach if the distribution amplitudes for $\eta$
and $\eta'$ mesons are taken close to the asymptotic one.
Particular attention is paid to the interpretation of the data
in terms of properly defined $\eta$-$\eta'$ mixing parameters.
I also comment on the use and misuse of interpolation formulas
for $P\gamma$ and $P\gamma^*$ transition
form factors.
{(\tt Talk presented at conference {\sc Photon'99}, Freiburg, May 1999)}
\end{abstract}

\maketitle

\section{Introduction}
\begin{fmffile}{ppic}

The hard exclusive production of 
pseudoscalar mesons $P=\pi^0,\eta^{\phantom.},\eta',\ldots$
in two-photon reactions
provides an important process to test our understanding of QCD and to
determine the properties of the pseudoscalar mesons.
The calculation of the related transition form factor 
\beq
&& \langle \gamma(q',\varepsilon)| J_{\mu}^{\rm elm} | P(p)\rangle 
\nonumber\\[0.2em] 
&=& i \, e_0^2 \, F_{P\gamma}(Q^2)\, 
 \epsilon_{\mu\alpha\beta\gamma} 
 \, p^\alpha \, q'{}^\beta \,\varepsilon^\gamma 
\eeq
at large virtualities
of (at least)
one of the two photons, $Q^2=(p-q')^2 \gg \Lambda_{\rm QCD}^2$,
is based on the factorization of 
the amplitude into a short- and
a long-distance part. The former can be calculated perturbatively
by considering the elementary scattering of photons and quarks.
In leading order this is a pure 
QED process, and thus uncertainties related to the
value of $\alpha_s$ only enter on the level
of QCD corrections. The long-distance part is expressed
in terms of process-independent
light-cone wave functions (LCWFs) of $q\bar q$- or eventually
higher Fock states in the meson. The LCWFs 
parametrize the non-perturbative
features related to, for instance, confinement.
In the asymptotic limit, $Q^2 \to \infty$, the behavior of the 
wave functions is known from the QCD evolution 
equation~\cite{BrLe80,Efremov:1980qk}. For the  
$\pi\gamma$ transition form factor this leads to the famous prediction
\beq
	\lim_{Q^2 \to \infty} Q^2 \, F_{\pi\gamma}(Q^2) &=&\sqrt2
  \, f_\pi \ .
\label{limit}
\eeq
The recent data from CLEO~\cite{CLEO97} at momentum transfer
up to $15$~GeV$^2$ are still 10-15\% below that limit, indicating
that corrections to Eq.~(\ref{limit}) have to be taken into
account. 
In the standard hard-scattering approach (sHSA), see e.g.\
Ref.~\cite{Brodsky:1997ia},
one uses the collinear approximation (i.e.\ one neglects the
transverse momenta of the partons) but includes radiative
gluon corrections in fixed order perturbation theory.
In the modified version of the hard-scattering approach (mHSA)
also the effects connected to the
transverse degrees of freedom
are taken into account.
In addition to the intrinsic transverse momenta
a resummation of radiative gluon
corrections accumulated in a Sudakov factor is 
included. The HSA gives a good description of the
$\pi\gamma$ transition form factor at moderate values
of $Q^2$ and indicates that the
pion wave function is close to its asymptotic form
already at small factorization scales,
for details
see e.g.\ Refs.~\cite{Brodsky:1997ia,KrRa96,MuRa97} and references therein.

In this talk I will concentrate on the mHSA calculation of the
$\eta\gamma$ and $\eta'\gamma$ transition form factors~\cite{FeKr97b}.
Using the $\pi\gamma$ form factor as a case of reference, one obtains
valuable information on the $\eta$ and $\eta'$ properties.
Especially
the $\eta$-$\eta'$ mixing parameters perfectly fit into
the improved theoretical and phenomenological pattern which has
been established recently in
Refs.~\cite{Leutwyler97,FeKrSt98}.

\section{MHSA calculation}

Starting point of the calculation is a Fock-state expansion
of the $\eta$ and $\eta'$ mesons 
\beq
|P\rangle
&=&  \sum_{a=8,0} \, { \psi_P^a(x,\vec k_\perp{})} \, 
    { |\bar q \, \frac{\lambda^a}{\sqrt2} \,  q\rangle} + 
\ldots
\label{Fock}
\eeq
Here and in the following $a=8(0)$ indicate flavor octet(singlet)
quantum numbers and $P=\eta,\eta'$. The dots stand for higher Fock
states with additional gluons or quark-antiquark pairs. Also a
pure two-gluon component $|gg\rangle$ can, in principle, contribute.
In Eq.~(\ref{Fock}) the non-perturbative information is encoded
in the LCWFs $\psi_P^a(x,\vec k_\perp{})$. 
Here $x$ denotes the ratio
of light-cone-plus components of quark
and  meson momenta. The transverse momentum of the quark
is denoted by $\vec k_\perp{}_i$. Due to momentum conservation
the anti-quark carries the momentum fraction $(1-x)$ and the
transverse momentum $-\vec k_\perp$.

We remind the reader that, in principle,
the parton distributions (familiar
from inclusive reactions)
can  be obtained from an infinite sum
over all $n$-particle
Fock states by taking the
squares of LCWFs and integrating over
all transverse momenta and all but one momentum fraction $x$ of
the struck parton~\cite{BrLe80}
\beq
q_i(x) &=&
\sum_n \int [dx] \, [d^2k_\perp] \, |\psi_n|^2 \, \delta(x_i-x) \ .
\eeq
It can be shown on quite general grounds that the lowest Fock states
dominate the parton distributions
at large values of $x$. 
In practice this feature can be exploited as a cross-check
for parametrizations of LCWFs, see e.g.\ Ref.~\cite{Diehl:1998kh}.
In parallel,
higher Fock states are suppressed by $1/Q^2$ in
exclusive reactions, like in the $P\gamma$ transition form factor. 
In what follows we 
therefore only consider the $|q\bar q\rangle$ Fock states.

As an ansatz for the LCWFs  we use
the same decomposition that has been 
proven successful for the pion~\cite{KrRa96}, i.e.\
\beq
  \Psi_P^a(x,\vec k) &=&
  \frac{f_P^a}{2\sqrt 6} \,
     {\phi_P^a(x,\mu)} \,
	\frac{16 \pi^2 (a_P^a)^2}{x \, (1-x)} \times \nonumber\\[0.2em] &&
        \, \exp\left[-{(a_P^a)^2} \frac{k_\perp^2}{x (1-x)}
			\right] \ .
\label{eq:psi}
\eeq
Here $\phi_P^a(x,\mu)$ is the 
distribution amplitude
for which we assume
the asymptotic form $\phi_{\rm AS}(x)=6 x (1-x)$. 
For the transverse size parameters $a_P^a$ we  
employ a single value taken from
the pion case. It is fixed by the chiral anomaly
prediction for $\pi^0\to\gamma\gamma$~\cite{BHL}, 
leading to $a_\pi=0.86$~GeV$^{-1}$.
We will see, that these assumptions are sufficient to explain
the experimental results for the $P\gamma$ transition form factors. 
Of course, they could be relaxed if more precise data
were available.

Finally, $f_P^a$ are the octet or singlet decay constants defined
as ($f_\pi=131~$MeV)
\beq
  \langle 0 | J_{\mu 5}^a | P(p)\rangle &=& i \, f_P^a \, p_\mu \ .
\label{fPadef}
\eeq
They play a distinguished role since, like in the pion case~(\ref{limit}),
they  determine the asymptotic limit of the $\eta\gamma$ and 
$\eta'\gamma$ transition form factors.
Due to the mixing in the $\eta$-$\eta'$ system both, $\eta$ and $\eta'$
mesons have octet and singlet components. The four decay constants,
resulting from the definition~(\ref{fPadef}), can
conveniently be parametrized as~\cite{Leutwyler97}
\beq
 \left(\matrix{f_\eta^8 & f_\eta^0\vspace{0.2em}\cr f_{\eta'}^8 & f_{\eta'}^0}
	\right) &=& \left( \matrix{ 
		f_8 \, \cos\theta_8 & - f_0 \, \sin\theta_0\vspace{0.2em} \cr
                f_8 \, \sin\theta_8 & \phantom{-} f_0 \, \cos\theta_0}
		\right) \ .
\eeq
For the numerical calculation we follow the phenomenological 
analysis in Ref.~\cite{FeKrSt98} and take
\beq
 f_8 \simeq 1.26\, f_\pi && \theta_8 \simeq -21.2^\circ \cr
          f_0 \simeq 1.17\, f_\pi && \theta_0 \simeq -\phantom{1}9.2^\circ
\ .
\label{80value}
\eeq
It is to be stressed that the two angles $\theta_8$ and $\theta_0$
are different. In $\chi$PT this difference is
induced by the
same $SU(3)_F$ breaking parameter $L_5$ which is 
responsible for the difference of kaon and pion decay constants
\beq
 \theta_8-\theta_0 \propto{L_5} \propto 1-f_K/f_\pi \ .
\eeq
For details see Ref.~\cite{Leutwyler97}.
We remind the reader that in $\chi$PT the decay constants 
relate the bare octet or singlet fields  with
the physical ones via
\beq
  P(x) &=& \sum_a f_P^a \, \varphi^a(x) \ .
\eeq
For $\theta_8\neq \theta_0$ this is not
a simple rotation, in contrast to our naive intuition,
\beq
{\left(\matrix{\eta \cr \eta'}\right)}
            \neq \left(\matrix{ \cos\theta & -\sin\theta  \cr
                                \sin\theta & \phantom{-}\cos\theta} \right)
\,
\left(\matrix{f_8 \, \varphi_8 \cr f_0 \, \varphi_0}\right)
\ .
\eeq

\begin{figure*}[htb]
\vspace{9pt}
\centerline{
\epsfclipon
a)\psfig{file=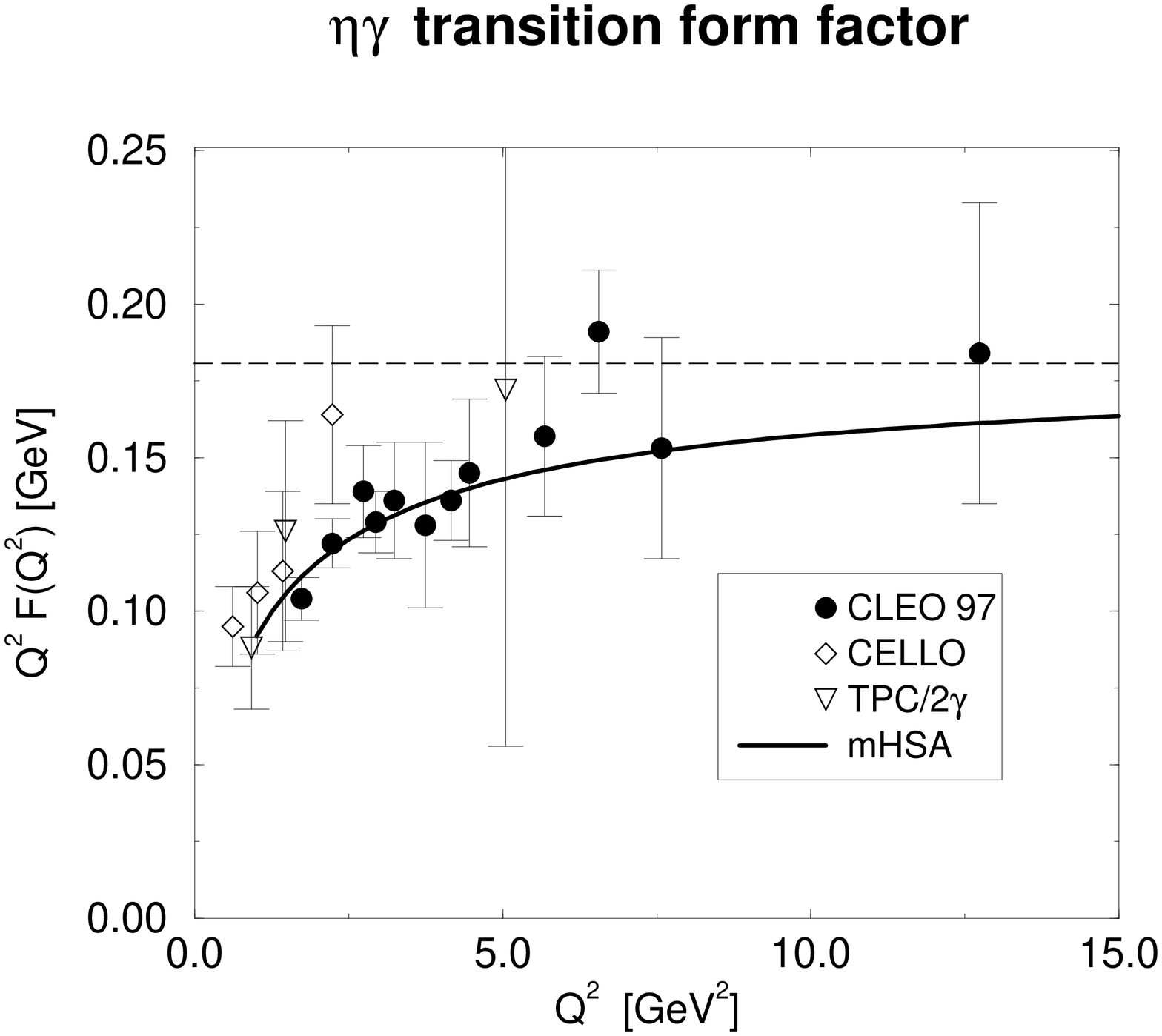,bb=90 20 590 630,width=5.05cm,angle=-90} \ \
b)\psfig{file=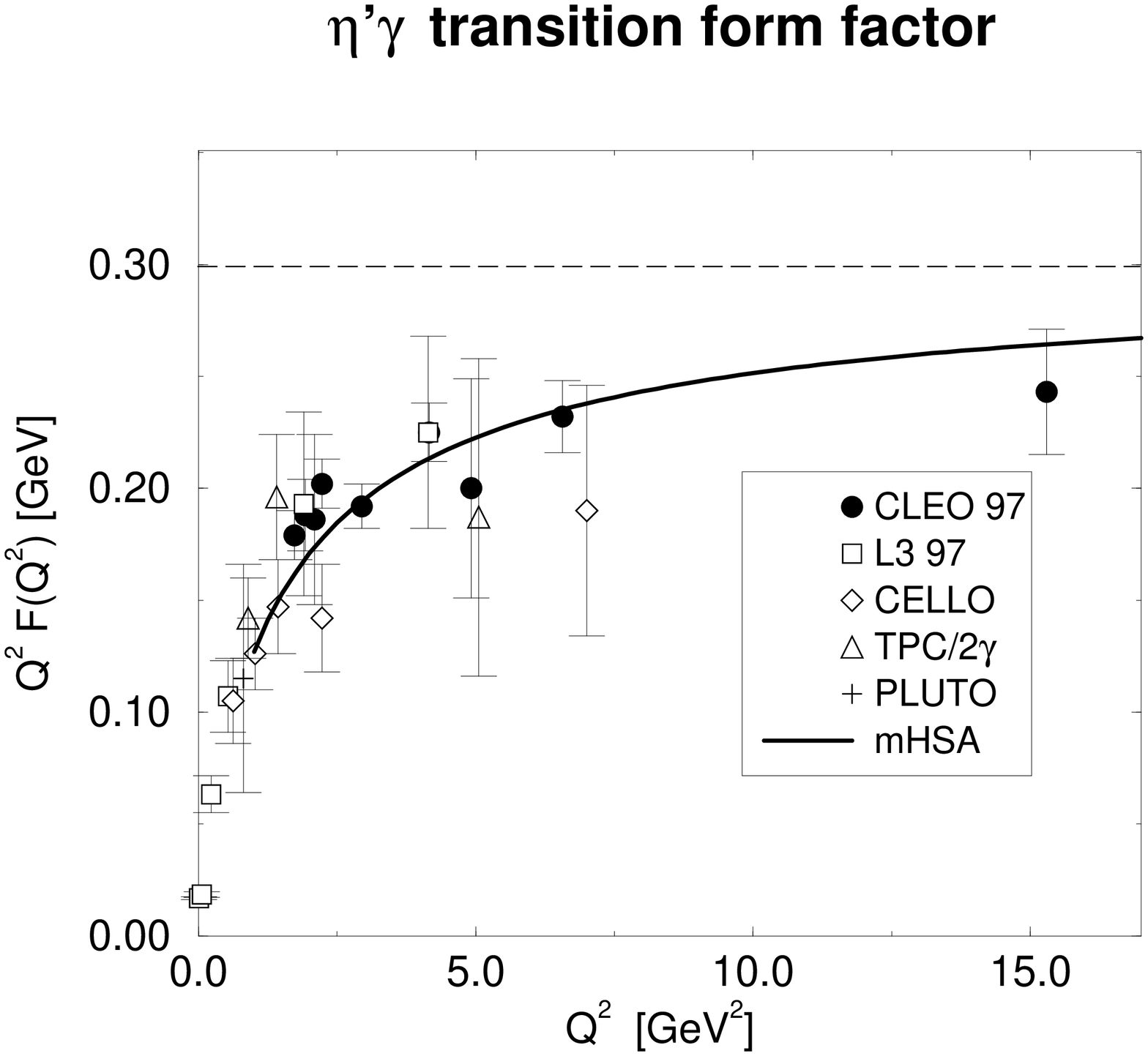,bb=90 20 590 630,width=5.05cm,angle=-90}
}
\caption{a) The $\eta\gamma$ transition form-factor, and b) 
the $\eta'\gamma$ transition form factor, each multiplied by $Q^2$.
The dashed lines indicate the asymptotic limit~(\ref{eq:etalimit}).
The solid line is the mHSA calculation. Data are taken from
Refs.~\cite{CLEO97,L397,CELLO91,TPC90,PLUTO}.}
\label{fig:result}
\end{figure*}

\begin{figure}[htb]
\vspace{9pt}
\centerline{
\unitlength=0.65cm
\parbox[c]{3.2\unitlength}{\fmfframe(0.1,0.1)(0.1,0.1){
\begin{fmfgraph}(3,2)
\fmfstraight 
\fmfleft{g1,g2}\fmfright{q1,q2}
\fmf{photon}{g1,v1}
\fmf{photon}{g2,v2} 
\fmf{fermion}{q1,v1,v2,q2}
\end{fmfgraph}}}
\hskip3em
\unitlength=0.65cm
\parbox[c]{3.2\unitlength}{\fmfframe(0.1,0.1)(0.1,0.1){
\begin{fmfgraph}(3,2)
\fmfstraight
\fmfleft{g1,g2}\fmfright{q1,q2}
\fmf{phantom}{g1,v1}
\fmf{phantom}{g2,v2} 
\fmf{fermion}{q1,v1,v2,q2}
\fmffreeze
\fmf{photon}{g2,v1}
\fmf{photon}{g1,v2} 
\end{fmfgraph}}}
}
\caption{Lowest order Feynman graphs contributing to the hard-scattering
	amplitude $T_H$.}
\label{fig:TH}
\end{figure}
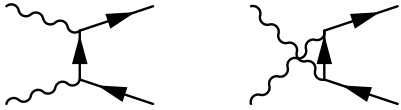

The $P\gamma$ transition form factors can now be calculated from
the following convolution formula
\beq
 F_{P\gamma}(Q^2)
&=& 
\sum_a \int dx \, \frac{d^2
b}{4\pi} \, {\hat
T_H^a(x,b,Q^2)} \times
\nonumber\\[0.1em]
&& \
{\exp[ - \mathcal S]}
\, {\hat \psi_P^a(x,b)} \ .
\label{eq:convol}
\eeq
A Fourier transformation to transverse configuration space,
$\vec k_\perp \to \vec b$, has been performed which is indicated
by the hat over $T_H^a$ and $\Psi_P^a$.
The hard-scattering amplitude is calculated
from the Feynman-graphs in Fig.~\ref{fig:TH},
\beq
T_H^a &=& \frac{4 \sqrt6 \, C_a}{x Q^2 + \vec k_\perp^2 + x (1-x) M_P^2}
\label{eq:TH}
\eeq
with $C_8=(e_u^2+e_d^2-2 e_s^2)/\sqrt6$ and $C_0=(e_u^2+e_d^2+e_s^2)/\sqrt3$
denoting charge factors. 
The term $\exp[-{\mathcal S}]$
denotes the Sudakov factor (its explicit form can be found in
Ref.~\cite{DaJaKr95}).
The
transverse separation $b$ of the two quarks inside the meson
provides an intrinsic definition of the (gliding) factorization scale
$\mu_F=1/b$ which sets the interface between true non-perturbative
soft gluon contributions -- still contained in the hadronic
wave functions -- and perturbative ones accounted for by the
Sudakov factor.

The integral in Eq.~(\ref{eq:convol}) can be calculated numerically
and leads to the result shown in Fig.~\ref{fig:result}.
One observes a perfect description of the experimental data for
$Q^2 \geq 1$~GeV$^2$. We also have shown the asymptotic limit
of the transition form factors which follows from Eqs.~(\ref{eq:psi}),
(\ref{eq:convol}) and (\ref{eq:TH})
\beq
	\lim_{Q^2 \to \infty} Q^2 \, F_{P\gamma}(Q^2) &=&
 \sum_a 6 \, C_a \, f_P^a \ .
\label{eq:etalimit}
\eeq
Using the parameters~(\ref{80value}),
one obtains for the r.h.s.\ of Eq.~(\ref{eq:etalimit}) the
values $181$~MeV and
$299$~MeV for $\eta$ and $\eta'$ mesons, respectively.
This has to be compared with the value  $185$~MeV
for the $\pi^0$, see Eq.~(\ref{limit}).
Note that the approximate equality of the
asymptotic limit for the $\pi^0\gamma$ and $\eta\gamma$ 
transition form factor is totally accidental and has nothing
to do with $SU(3)_F$ flavor symmetry within the pseudoscalar octet
(which is, of course, broken here by the electric charges).

\section{Interpretation of the data}

One of the
conclusions drawn at the {\sc Photon'97} conference
on the basis of the CLEO data for the $P\gamma$ transition
form factors was 
that the $\eta'$ meson behaves differently
than the other light pseudoscalars~\cite{Savinov:1997kc}.
Although, at first glance, this statement is not surprising since
the $\eta'$ is known to be effected by the $U(1)_A$ anomaly, it
is in obvious contradiction to the results presented in the
previous section where a good description of the
data has been achieved with similar 
LCWFs for $\pi^0$, $\eta$ and $\eta'$ mesons.
The resolution of the apparent contradiction is connected with
the treatment of the data at large and small values of $Q^2$.

\begin{figure}[hbt]
\vspace{9pt}
\centerline{
\epsfclipon
\psfig{file=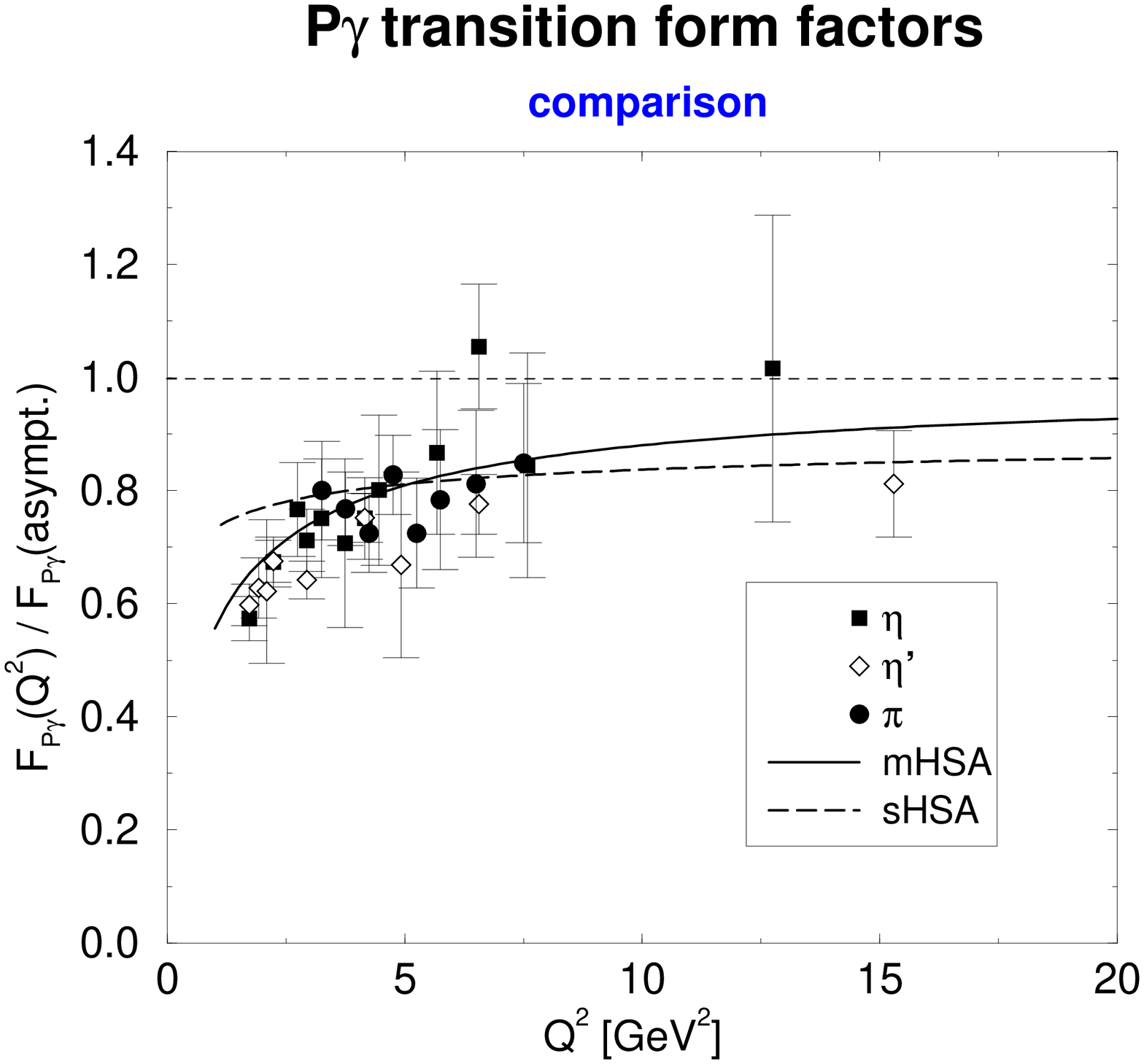,bb=95 30 590 645,width=5.3cm,angle=-90} \ \
}
\caption{Comparison of $P\gamma$ transition form factors for
$P=\pi^0,\eta,\eta'$. Data are taken from CLEO~\cite{CLEO97} and
divided by the asymptotic behavior
(\ref{limit}) and (\ref{eq:etalimit}).
The solid line is the mHSA prediction~(\ref{eq:convol}); the dashed line the
sHSA prediction~(\ref{shsa}), the dotted line the asymptotic limit.
Meson masses have been neglected in this plot.}
\label{compfig}
\end{figure}

First, let us plot again the large-$Q^2$ data from CLEO 
for $\pi^0$, $\eta$ and $\eta'$ but now divided by
their asymptotic behavior, Eqs.~(\ref{limit}) and 
(\ref{eq:etalimit}),
respectively, see Fig.~\ref{compfig}.
Basically, within the errors, the data for the three
different mesons fall on top of each other. For comparison, we have
again included the result of the mHSA approach~(\ref{eq:convol}) --
with meson masses neglected in the hard-scattering amplitude~(\ref{eq:TH}) --
and also the sHSA prediction~\cite{Brodsky:1997ia}
\beq
  Q^2 \, F_{\pi\gamma}(Q^2) &=&
\sqrt2 \,f_\pi \ \left(1 - \frac 53 \, \frac{\alpha_V(e^{-3/2} Q)}{\pi} \right)\ . \nonumber\\[0.15em] &&	
\label{shsa}
\eeq
In both cases we have employed
the asymptotic distribution amplitude.
Note that, following Ref.~\cite{Brodsky:1997ia},
in Eq.~(\ref{shsa}) a {\it freezing} strong coupling constant
$\alpha_V$ is utilized at rather small renormalization scales. 
Both, the sHSA and the mHSA, describe the data reasonably well.

As discussed above, the $P\gamma$ transition form factor at large values
of momentum transfer is dominated by the $|q\bar q\rangle$ Fock states.
We therefore have to conclude that $\pi^0$, $\eta'$ and $\eta'$ mesons behave
similarly in hard exclusive reactions due to the fact that the
anomalous character of the $\eta'$ meson does not show up in the 
$|q\bar q\rangle$ Fock states.

On the other hand, in experimental analyses one often 
performs a fit to the data which is based on a simple 
pole ansatz
\beq
 F_{P\gamma}(Q^2) &\propto & \frac{ 1}{1 + Q^2/\Lambda_P^2} \ .
\label{polefit}
\eeq
The fit includes data at both, low and 
high values of momentum transfer. The effective pole masses 
found from e.g.\ the CLEO analysis~\cite{Savinov:1997kc} are
$\Lambda_\pi=776 \pm 22$~MeV,
$\Lambda_\eta=774 \pm 29$~MeV,
and
$\Lambda_{\eta'}=859 \pm 28$~MeV.
Apparently, the $\eta'$ behaves differently in that
kind of analyses. But is there any deeper physical meaning in the
parameters $\Lambda_P$?
To answer this question it is necessary to consider interpolation
formulas for the $P\gamma$ transition form factors which are usually
taken as a justification of Eq.~(\ref{polefit}).
For the pion such a simple formula has been proposed by 
Brodsky/Lepage~\cite{BrLe80}
\beq
F_{\pi\gamma}^{BL}(Q^2) & = &
\frac{\sqrt2 \, f_\pi}{Q^2 + 4\pi^2 f^2_\pi}
 \ .
\label{BLint}
\eeq
Obviously, it has the correct asymptotic limit~(\ref{limit}).
Furthermore for $Q^2 \to 0$ it coincides with the chiral anomaly
 prediction. 
Eqs.~(\ref{BLint}) and (\ref{polefit}) 
also {\em happen} to have a similar form as
the vector dominance model (VDM).
One also has the approximate
equality $\Lambda_\pi \simeq 2 \pi f_\pi \simeq M_{\rho,\omega}$,
but there is no reason for these
relations to be exact.

Things become more involved if we consider the $\eta\gamma$ and $\eta'\gamma$
transition form factors. 
Here both, the asymptotic limit~(\ref{eq:etalimit}) as well
as the chiral anomaly prediction for $Q^2=0$~\cite{Leutwyler97,FeKr97b}
are given by a linear combination of two terms, arising from
the mixing of octet and singlet contributions.
For $\theta_8\neq \theta_0$
the pre-factors in the limit $Q^2\to 0$ and $Q^2\to \infty$ 
are {\em different},
and one does not find a simple interpolation formula.
It has been observed~\cite{FeKrSt98} that the decay constants which
enter both the theoretical predictions, for $Q^2 =0$ and $Q^2 \to \infty$,
have (at least approximately)
a much simpler decomposition in the quark flavor basis, 
\beq
 \left(\matrix{f_\eta^q & f_\eta^s \vspace{0.2em}\cr f_{\eta'}^q & f_{\eta'}^s}
	\right) &=& \left( \matrix{ 
		f_q \, \cos\phi & - f_s \, \sin\phi \vspace{0.2em} \cr
                f_q \, \sin\phi & \phantom{-} f_s \, \cos\phi}
		\right) \ .
\label{phi}
\eeq
Here the index $q$ denotes the flavor combination $(u\bar u+d\bar d)/\sqrt2$
and $s$ stands for $s\bar s$. 
The parameters in Eq.~(\ref{80value}) correspond to
$\phi=39.3^\circ$, $f_q=1.07 \, f_\pi$, $f_s = 1.36 \, f_\pi$.
In the quark flavor basis it is sufficient to consider only one single
mixing angle $\phi$.
This is equivalent to neglecting
all true OZI-rule violating contributions
while topological effects connected to the $U(1)_A$ anomaly
and mixing are kept.
In this scheme
 interpolation formulas can be obtained~\cite{FeKr97b} as a linear combination
of two individual terms which resemble the Brodsky/Lepage formula
\beq
F_{P\gamma}^{\rm int}(Q^2) & = &
\sum_{i=q,s} 
\frac{6 \, C_i \, f_P^i}{Q^2 + 4 \pi^2 \, f_i^2}
\ , 
\qquad (P=\eta,\eta') \nonumber \\[0.2em] &&
\label{etainterpolqs}
\eeq
with $C_q=5/9\sqrt2$ and $C_s=1/9$.
Again we have an approximate relation to VDM,
since $M_{\rho,\omega} \simeq 2 \pi f_q$ and $M_\phi \simeq 2 \pi f_s$.
A simple
connection to the experimental fit parameters $\Lambda_\eta$ and 
$\Lambda_{\eta'}$ can, however, not be found. We may expect that
the values of $\Lambda_\eta$ and $\Lambda_{\eta'}$ lie 
somehow between $M_\rho$ and $M_\phi$. In fact, since the charge
factor in front of the strange quark contribution in
Eq.~(\ref{etainterpolqs}) is smaller than the one for the light
quark contribution the values of $\Lambda_P$ should be closer to
the $\rho$ mass. Furthermore, for $\phi=39.3^\circ$ the strange
quark component of the $\eta'$ meson is larger than the one for
$\eta$; thus one should have $\Lambda_{\eta'} > \Lambda_{\eta}$.
This is in line with the experimental findings.

Coming back to our question from above, we have to consider
the pole-parameters $\Lambda_P$ as an effective way to compare
results of different experiments (as long as they are performed at
similar values of $Q^2$). They have no deeper theoretical meaning,
although they show some qualitative
similarities with VDM.
In particular, the values of $\Lambda_P$ cannot be used as a
measurement of the decay constants $f_\pi$ and $f_P^a$.

Furthermore, interpolation formulas like Eqs.~(\ref{BLint}),
(\ref{etainterpolqs}) have
to be used with some care: Compared to the mHSA prediction they
tend to overestimate the transition form factor at intermediate
momentum transfer~\cite{FeKr97b}. 
Nevertheless, they can be used to obtain a
rough estimate, e.g.\ for the implementation
of the transition form factors in Monte Carlo generators~\cite{Gulik}.

If both photons are
taken far off-shell one should also keep in mind that the
naive VDM-prediction leads to the wrong asymptotic behavior
($\propto 1/Q_1^2 Q_2^2$).
The correct limit is obtained from a generalization
of Eq.~(\ref{eq:etalimit}) 
\beq
&& \lim_{Q_1^2,Q_2^2 \to \infty}
\,  F_{P\gamma^*}(Q_1^2,Q_2^2)\nonumber\\[0.2em]
& = & 
  \sum_{i=q,s}   6 \, C_i \, f_P^i \,
\int_0^1 dx \, \frac{2 x (1-x)}{
	x Q_1^2 + (1-x)  Q_2^2} \ . \nonumber \\ && 
\label{gammastarlimit}
\eeq
Starting from Eq.~(\ref{gammastarlimit}), again interpolation
formulas for the general case $Q_1^2,Q_2^2 \neq 0$ can be
constructed. A simple choice, which leads to results that are very
close to the mHSA calculation, is to perform in
Eq.~(\ref{gammastarlimit}) the replacement, 
$$x Q_1^2 + (1-x)  Q_2^2 \to x Q_1^2 + (1-x)  Q_2^2 + 4\pi^2 f_i^2/3
\ .$$
The $x$-integration in Eq.~(\ref{gammastarlimit}) can be performed
analytically, but the result looks rather complicated.
For a further discussion and alternative choices of
interpolation formulas see also 
the work of  Ong~\cite{Ong:1995gs}
which has been contributed to the {\sc Photon'97}
conference.
In any case, for the interpolation formulas for $P\gamma^*$
transition form factors
the same limitations apply as for the approximations~(\ref{BLint})
and (\ref{etainterpolqs}).

\section{Conclusions}

The experimental data
for $\eta\gamma$ and $\eta'\gamma$ transition form factors
at large momentum transfer are well described by the hard-scattering
approach. Two important features of $\eta$ and $\eta'$ mesons can
be read off from the data: i) The distribution amplitudes of the
quark-antiquark Fock states of $\eta$ and $\eta'$ mesons are similar to
the pion one and close to the asymptotic form already at low scales.
This means that the $U(1)_A$ anomaly does not show up  in the
shape of the LCWFs of the valence Fock states.
A similar conclusion has been drawn within a somewhat different
framework in Ref.~\cite{Anisovich97}.
ii) The values of the decay constants
in the $\eta$-$\eta'$ system, which have been determined from
a detailed phenomenological analysis elsewhere~\cite{FeKrSt98}, 
are confirmed. This underlines
the necessity of using the general mixing scheme presented in
Ref.~\cite{Leutwyler97} whereas a description in terms of
a single octet-singlet mixing angle is disfavored.

The interpretation of the experimental data by means of
effective pole masses based on
simplified interpolation formulas can be very misleading
and has to be considered
with great care. At most, such a procedure
can be used to compare different experiments, provided that the
data is taken within similar ranges of momentum transfer.
The pole masses do not at all
provide a process-independent definition of \lq{}decay constants\rq{}.

The results can easily be generalized to the case
where both photons are off-shell. Experimental data for $P\gamma^*$
transition form factors is, however, not available yet.


\end{fmffile}

\end{document}